\documentclass[twocolumn]{aastex631}

\newcommand{\kms}{{\rm km~s^{-1}}}

\newcommand{\simgt}{\,\rlap{\lower 3.5 pt \hbox{$\mathchar \sim$}} \raise
1pt \hbox {$>$}\,}
\newcommand{\simlt}{\,\rlap{\lower 3.5 pt \hbox{$\mathchar \sim$}} \raise
1pt \hbox {$<$}\,}

\newcommand{\hb}{${\rm H\beta}$}
\newcommand{\hg}{${\rm H\gamma}$}
\newcommand{\hd}{${\rm H\delta}$}

\newcommand{\oiii}{[\textrm{O}~\textsc{iii}]}

\newcommand{\Z}{{\rm [Z/H]}}

\submitjournal{ApJ Letters}
\accepted{October 8, 2023}

\shorttitle{Puzzling Nature of JD1 revealed by JWST observations}
\shortauthors{Stiavelli et al.}
\graphicspath{{./}{figures/}}
\begin{document}

\title{ 
The puzzling properties of the MACS1149-JD1 galaxy at z=9.11
}

\author[0000-0001-9935-6047]{Massimo Stiavelli}
\affiliation{Space Telescope Science Institute, 3700 San Martin Drive, Baltimore, MD 21218, USA}

\author[0000-0002-8512-1404]{Takahiro Morishita}
\affiliation{IPAC, California Institute of Technology, MC 314-6, 1200 E. California Boulevard, Pasadena, CA 91125, USA}

\author[0000-0003-1564-3802]{Marco Chiaberge}
\affiliation{Space Telescope Science Institute for the European Space Agency (ESA), ESA Office, 3700 San Martin Drive, Baltimore, MD 21218, USA}
\affiliation{The William H. Miller III Department of Physics and Astronomy, Johns Hopkins University, Baltimore, MD 21218, USA}

\author[0000-0002-5926-7143]{Claudio Grillo}
\affiliation{Dipartimento di Fisica, Università degli Studi di Milano, Via Celoria 16, I-20133 Milano, Italy}
\affiliation{INAF - IASF Milano, via A. Corti 12, I-20133 Milano, Italy}

\author[0000-0003-4570-3159]{Nicha Leethochawalit}
\affiliation{National Astronomical Research Institute of Thailand (NARIT), Mae Rim, Chiang Mai, 50180, Thailand}

\author[0000-0002-6813-0632]{Piero Rosati}
\affiliation{INAF - OAS, Osservatorio di Astrofisica e Scienza dello Spazio di Bologna, via Gobetti 93/3, I-40129 Bologna, Italy}
\affiliation{Dipartimento di Fisica e Scienze della Terra, Università degli Studi di Ferrara, Via Saragat 1, I-44122 Ferrara, Italy}

\author[0000-0003-2497-6334]{Stefan Schuldt}
\affiliation{Dipartimento di Fisica, Università degli Studi di Milano, Via Celoria 16, I-20133 Milano, Italy}
\affiliation{INAF - IASF Milano, via A. Corti 12, I-20133 Milano, Italy}

\author[0000-0001-9391-305X]{Michele Trenti}
\affiliation{School of Physics, University of Melbourne, Parkville 3010, VIC, Australia}
\affiliation{ARC Centre of Excellence for All Sky Astrophysics in 3 Dimensions (ASTRO 3D), Australia}

\author[0000-0002-8460-0390]{Tommaso Treu}
\affiliation{Department of Physics and Astronomy, University of California, Los Angeles, 430 Portola Plaza, Los Angeles, CA 90095, USA}



\begin{abstract}
We analyze new JWST NIRCam and NIRSpec data on the redshift 9.11 galaxy MACS1149-JD1. Our NIRCam imaging data reveal that JD1 comprises three spatially distinct components. Our spectroscopic data indicate that JD1 appears dust-free but is already enriched, $12 + \log {\rm (O/H) } = 7.90 ^{+0.04}_{-0.05}$. We also find that the Carbon and Neon abundances in JD1 are below the solar abundance ratio. Particularly the Carbon under-abundance is suggestive of recent star formation where Type~II supernovae have already enriched the ISM in Oxygen but intermediate mass stars have not yet enriched the ISM in Carbon. A recent burst of star formation is also revealed by the star formation history derived from NIRCam photometry. Our data do not reveal the presence of a significant amount of old populations, resulting in a factor of $\sim7\times$ smaller stellar mass than previous estimates. Thus, our data support the view that JD1 is a young galaxy.
\end{abstract}

\keywords{}
%Classical Novae (251) --- Ultraviolet astronomy(1736) --- History of astronomy(1868) --- Interdisciplinary astronomy(804)}


\textbf{}\section{Introduction} \label{sec:intro}

The high sensitivity of the NIRSpec spectrometer onboard JWST \citep[][]{Boeker2023} has enabled us to study emission line diagnostics of galaxies at redshift $z>8$ with unprecedented details.
%DOI? %https://mast.stsci.edu/portal/Mashup/Clients/Mast/Portal.html?searchQuery=%7B%22service%22%3A%22CAOMDB%22%2C%22inputText%22%3A%22110.826424%20-73.4507599%22%2C%22paramsService%22%3A%22Mast.Caom.Cone%22%2C%22title%22%3A%22MAST%3A%20110.826424%20-73.4507599%22%2C%22columns%22%3A%22*%22%2C%22caomVersion%22%3Anull%7D
Already the ERO observations of SMACS J0723.3–7327 S04590 at $z=8.496$ \citep[][]{Curti2023, Heintz2023} enabled the first detection at high redshift of the auroral line [OIII]$\lambda4363$ and its use to derive an electron temperature $T_e \geq 24,000$\,K. The derived metallicity for this object is $12 + \log {\rm (O/H) } \sim 7$. Comparison with strong emission line diagnostics showed that this object was outside the validity range of most calibrations.
A broader look at a small sample of objects with auroral lines by \citet{Laseter2023, Sanders2023} finds that abundances derived from the $T_e$ method are generally not consistent with those from locally calibrated strong emission lines.

The galaxy MACS1149-JD1 \citep{Zheng2012} (hereafter JD1) was one of a handful of confirmed $\geq9$ galaxies known before the launch of JWST and, being substantially lensed, it was a credible candidate for searching for a potentially very low metallicty proto-galaxy. The high-redshift nature of MACS1149-JD1 was confirmed thanks to ALMA detection of [OIII]$\lambda88\micron$ \citep{Hashimoto2018,tokuoka2022}. 
%
%JD1 has been known to be relatively luminous, and massive, galaxy. 
The previous measurements by Spitzer indicate the presence of old stellar populations, making it a relatively massive system among other galaxies at similar redshifts. These exceptional properties of JD1 led us to make it as one of the prime candidates for deep spectroscopy with JWST.

It is worthwhile noticing that another bright, pre-JWST high-$z$ galaxy, GN-z11 \citep{Oesch2016}, while extremely interesting \citep[see e.g.][]{CameronGNz112023, Tacchella2023, BunkerGNz11, CharbonnelGNz112023A&A, Maiolino2023}, is considered to possibly host an AGN and is at a redshift where [OIII]$\lambda$5007 is beyond the NIRSpec sensitivity range and therefore makes it less likely to be able to carry out a direct metallicity measurement using the auroral lines. 
%\citet{Boyett2023} reported Gz9p3, another luminosu galaxy at redshift 9.3, that appears interacting and star forming. Unfortunately, the observations did not include \hb\ and \oiii\ and this makes it harder to carry out a comparable analysis. %comparable to the one we carry out here. 
As such, JD1 is currently one of a few luminous galaxies at $z>9$ that allow a reliable auroral line analysis.\footnote{\citet{Boyett2023} recently reported Gz9p3, another luminous galaxy at $z=9.3$, that appears interacting and starforming. Unfortunately, the observations did not cover the \hb\ and \oiii\ lines, preventing us from conducting a comparable analysis.}

In this paper, we present a comprehensive analysis of the properties of JD1 based on new observations by JWST. We describe our data on JD1 in Section \ref{sec:data}.  Section \ref{sec:sfrate} derives the H$\beta$ based star formation rate. Section \ref{sec:dust} is devoted to deriving an estimate for dust extinction and presence of dust as well as constraints on the electron density. Section \ref{sec:emlines} derives the metallicity of JD1 through the direct method based on auroral lines and discusses implications from the other lines. Section \ref{sec:COratio} discusses the non-solar abundance ratios for JD1. The star formation history and stellar mass are described in Section \ref{sec:sfh}. Section \ref{sec:disc} discusses our conclusions.

Throughout the paper, we adopt the AB magnitude system \citep{oke83,fukugita96}, cosmological parameters of $\Omega_m=0.3$, $\Omega_\Lambda=0.7$, $H_0=70\,\kms\, {\rm Mpc}^{-1}$, and the \citet{chabrier03} initial mass function. We express all measurements as a function of $\mu/10$ where relevant.

%%%%%%%%%%%%%%%%%%%%%

%%%%%%%%%%%%%%%%%%%%%
\section{Data and Analyses}\label{sec:data}
\subsection{NIRCam photometry}\label{sec:nircam}

NIRCam imaging observations (GTO \#1199) were executed on June 6 -- 8, with six filters configured (F090W, F115W, F150W, F200W, F277W, F356W, F444W), with $\sim1.1$\,hrs exposure each. We conduct photometry on the newly taken NIRCam images, along with the archival HST images that were originally taken in several HST programs \citep[CLASH, HFF, GLASS, BUFFALO;][]{postman12,treu15,lotz17,kelly18,steinhardt20}. 
We follow the same procedure presented in \citet{morishita23} for the image reduction and photometry. We hereby provide a high level description of our workflow: raw NIRCam images are reduced by using the official {\tt jwst} pipeline \citep[ver1.10.0, with context pmap \#1069][]{bushouse2023_7795697}, with several customized steps included to effectively remove artifacts and improve cosmic ray rejection. The final drizzled images are aligned to the WCS of GAIA. Source fluxes are calculated by using {\tt SExtractor} \citep{bertin96} in PSF-matched images to the psf of F444W, in an aperture of $r=0.\!''16$, reaching to $5\,\sigma$ limiting magnitude of $\sim 27.9$--28.8\,mag.
%The total exposure time of each filter image is $1.14$\,hrs.

\begin{figure*}
\centering
	\includegraphics[width=0.95\textwidth]{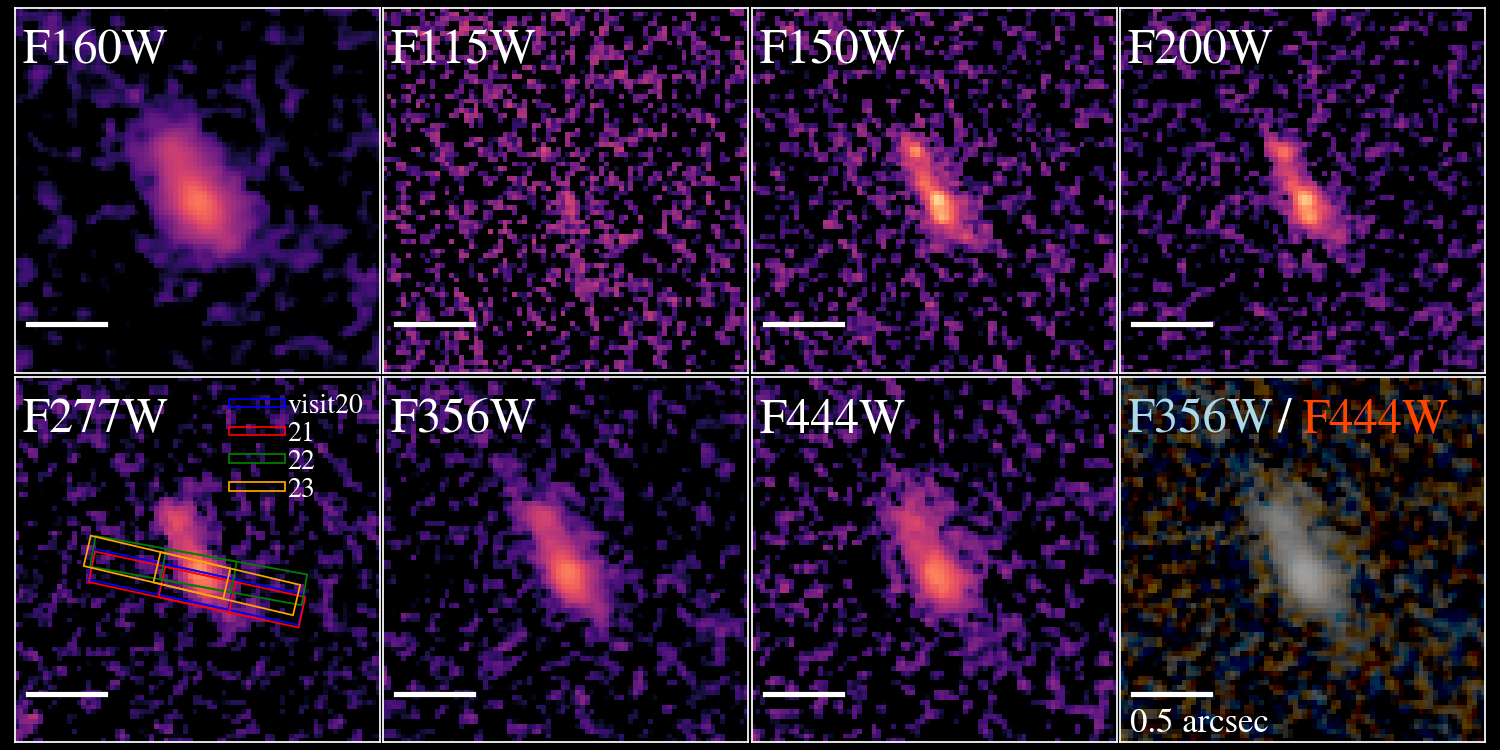}
	\caption{
    Stamp images of MACS1149-JD1 in HST F160W and NIRCam filters ($2.\!''4\times2.\!''4$). The MSA slits of the four visits (\#20--23) are overlaid in the F277W stamp image. The pseudo 2-color image using the F356W (blue) and F444W (red) filters, which corresponds to the Balmer break, is shown (right bottom).
    }
\label{fig:cs}
\end{figure*}

The NIRCam images of JD1 (see Figure \ref{fig:cs}) reveal the presence of three main components, two of which are included in our NIRSpec MSA slits (Sec.~\ref{sec:nirspec}). This means that geometry might play a significant role in the interpretation of the observations. However, we do not observe significant color differences between these components and therefore in our following analysis we will not consider them separately.

%%%%%%%%%%%%%%%%%%%%%
\subsection{NIRSpec MSA spectroscopy}\label{sec:nirspec}

Our NIRSpec observations were executed over four different visits. In the first two visits, \#20 and \#22 respectively for G235M and G395M, were taken at PA=257.766. In the second group of visits, the same grating pair was used, but with a slightly different PA, namely PA=259.660 for visit \#21/G235M and PA=256.766 for visit \#23/G395M. Slightly offset pointings ensure that roughly the same area of JD1 is covered by all exposures. 

We reduce the MSA spectra using {\tt msaexp}\footnote{https://github.com/gbrammer/msaexp} (ver0.6.13), following the procedure presented in \citet[][also, Morishita et al., in prep]{morishita23b}. The two-dimensional sky background is estimated by nodding the stacked spectrum for $6$\,pixels and then it is subtracted. The one-dimensional spectrum is optimally extracted \citep{horne86} by using the one-dimensional source profile derived from the 2-dimensional stacked spectrum as weight. For the extracted 1-dimensional spectrum, we fit each line of interest with a Gaussian, after subtracting the underlying continuum spectrum, estimated by scaling the best-fit SED template (see Sec.~\ref{sec:sfh}). We note that our flux measurements include a small correction for absorption in the Balmer series, which are inferred by the stellar template of the best-fit SED model. The correction is 1\% for \hb, 3\% for \hg, and 4\% for \hd. For each line, the total flux is estimated by integrating the flux over the wavelength range of 2$\times$FWHM, the latter derived from the Gaussian fit. For \oiii-doublet, we fix the ratio to 1:3 and adopt a single parameter for the widths of both lines. Lastly, we determined an aperture correction by scaling the difference between the continuum measured around individual spectral lines and the one inferred by the best-fit SED template. We find that this correction is well described by the slit-loss correction one derives for perfectly centered source \citep[][NIRSpec MOS Operations Slit Losses]{jdox} rescaled by the factor  $\sim2.36$ we find for \hb\ and \oiii\ .  This correction is applied to measured line fluxes.  However, most of our results are relatively unaffected by this wavelength dependent aperture correction. %The flux measurements presented below are scaled. 
The aperture correction allows us to apply measurements such as the Star Formation Rate (SFR) to the whole galaxy.
In what follows, when necessary we adopt the redshift $z=9.114$ obtained for the \hb+\oiii\ lines. 

We have checked that the line ratios obtained in all visits are consistent. Therefore, in the following we coadd the spectra and focus our analysis on the combined spectra. The measured line fluxes are reported in Table~\ref{tab:prop}.

% \citet{Tokuoka2022} found evidence for rotation in JD1. Our slits are misaligned with the position angle of the rotation pattern. As a consequence we do not detect significant rotation at our slit position angle.

%%%%%%%%%%%%%%%

%%%%%%%%%%%%%%%%%%%%%

%%%%%%%%%%%%%%%%%%%%%
\begin{figure*}
\centering
	\includegraphics[width=0.99\textwidth]{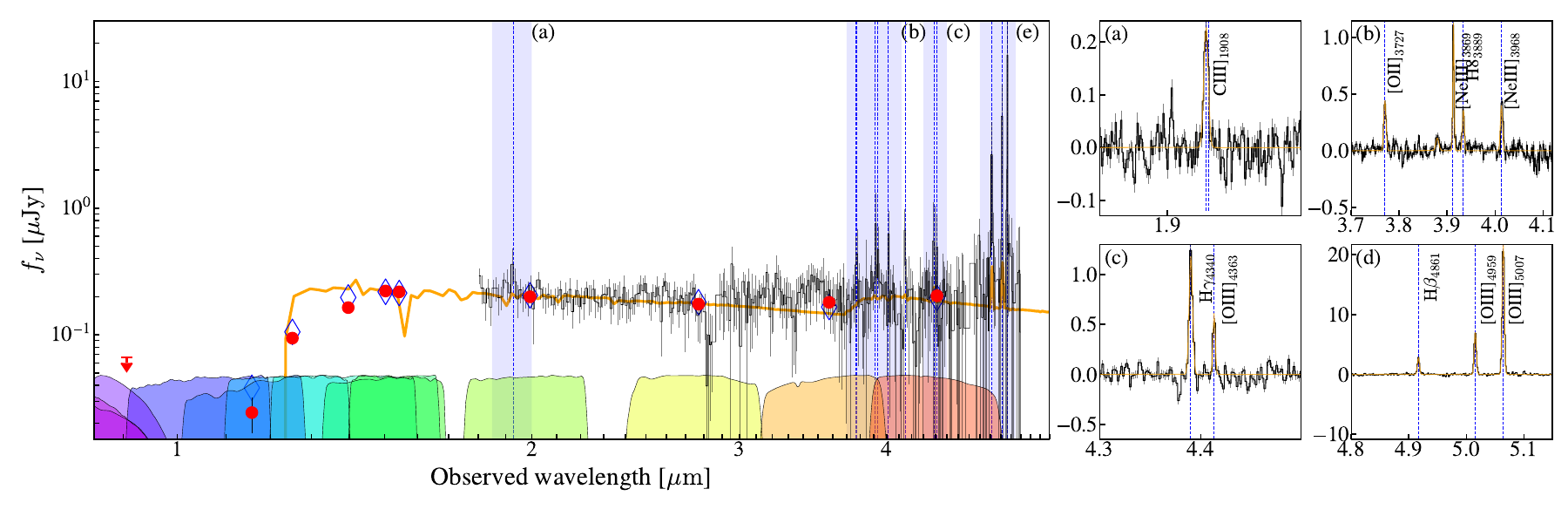}
	\caption{
    ($Left$): Spectral energy distribution of JD1 inferred by {\tt gsf} (orange solid line; Sec.~\ref{sec:sfh}). Observed NIRCam + HST photometric fluxes (red circles; inverted triangles for $2\,\sigma$-upper limits) and NIRSpec MSA G235M+G395M spectra (black lines) are included in the fit. Note that the observed fluxes are presented with $2\,\sigma$ uncertainties and not corrected for magnification. 
    ($Right$): Zoom-in plots of the continuum-subtracted spectra around strong emission lines. Fitted Gaussian profiles are shown (orange solid lines).
    %This was obtained by fixing metallicity to the direct value from the spectra and $A_V = 0.0$.
	}
\label{fig:sed}
\end{figure*}

\section{Star formation rate}\label{sec:sfrate}

The measured flux in \hb\ gives us an estimate of the star formation rate in JD1. Following the same approach as \citet{Heintz2023}, we assume the Case B ratio for $f_{\rm H\alpha / H\beta} = 2.80$ for $T_e = 1.6 \times 10^4$ K as derived from the direct analysis described in Section \ref{sec:emlines}. We derive the star formation rate as:

\begin{equation}
    {\rm SFR_{H\beta}} = 5.5 \times 10^{-42} L_{\rm H\beta} \times f_{\rm H\alpha / H\beta}\,{\rm (erg/s)},
\end{equation}
From the measured \hb\ flux of $3.7_{-0.1}^{+0.1} \times 10^{-18}$\,erg\,s$^{-1}$\,cm$^{-2}$ and assuming a gravitational magnification factor $\mu = 10$ \citep{Hashimoto2018}\footnote{\citet{Grillo2016} model predicts a median magnification of $12^{+11}_{-5}$ at the position of JD1, where the $\pm 1\sigma$ values are obtained by extracting 200 random sets of parameter values from the MCMC model chains. }, we obtain SFR$_{\rm H\beta} = 5.9_{-0.2}^{+0.2}\,(10/\mu)$\,M$_\odot$\,yr$^{-1}$. Note that this is the aperture-corrected value of the SFR  (Sec.~\ref{sec:nirspec}). The estimated \hb-based SFR is consistent with the one derived with \oiii\,88\,$\mu$m \citep[$4.2_{-1.1}^{+0.8}\,(10/\mu)\,{\rm M_\odot\,yr^{-1}}$;][]{Hashimoto2018}. Also, the estimate is roughly consistent with the one from our SED analysis, $\sim3.2\,(10/\mu)\,{\rm M_\odot\,yr^{-1}}$, derived from the rest-frame UV luminosity by following \citet[][]{morishita23c}.

Given the large uncertainties in the magnification values at the location of JD1 \citep[e.g.][]{Grillo2016,Finney2018} and the relative insensitivity of our results on the specific magnification we are expressing our results as a function of $\mu/10$ rather than adopting a specific value. 

%=============================

\section{Dust content and electron density}\label{sec:dust}

We have attempted to determine the presence of dust by looking at the Balmer decrement. We measure ${\rm H\gamma / H\beta} = 0.46 \pm 0.03$ as compared to the Case B recombination value of 0.47 for $T_e = 1.6 \times 10^4$\,K. This value would suggest a small $A_V=0.12$ but is also compatible with no dust. We also find ${\rm H\delta / H\beta} = 0.33 \pm 0.03$ as opposed to the Case B value of 0.26. The measured value is incompatible at 2.3+$\sigma$ with dust. These conclusions are supported also by analysing both pointings separately.
% Note also that we have not applied any correction for Balmer absorption. Such a correction would further increase the value of the ratio as the correction is relatively flat and tends to enhance higher order lines compared to lower order ones. 

\begin{figure}
\centering
	\includegraphics[trim={0 5cm 0 2.5cm},clip,width=0.52\textwidth]{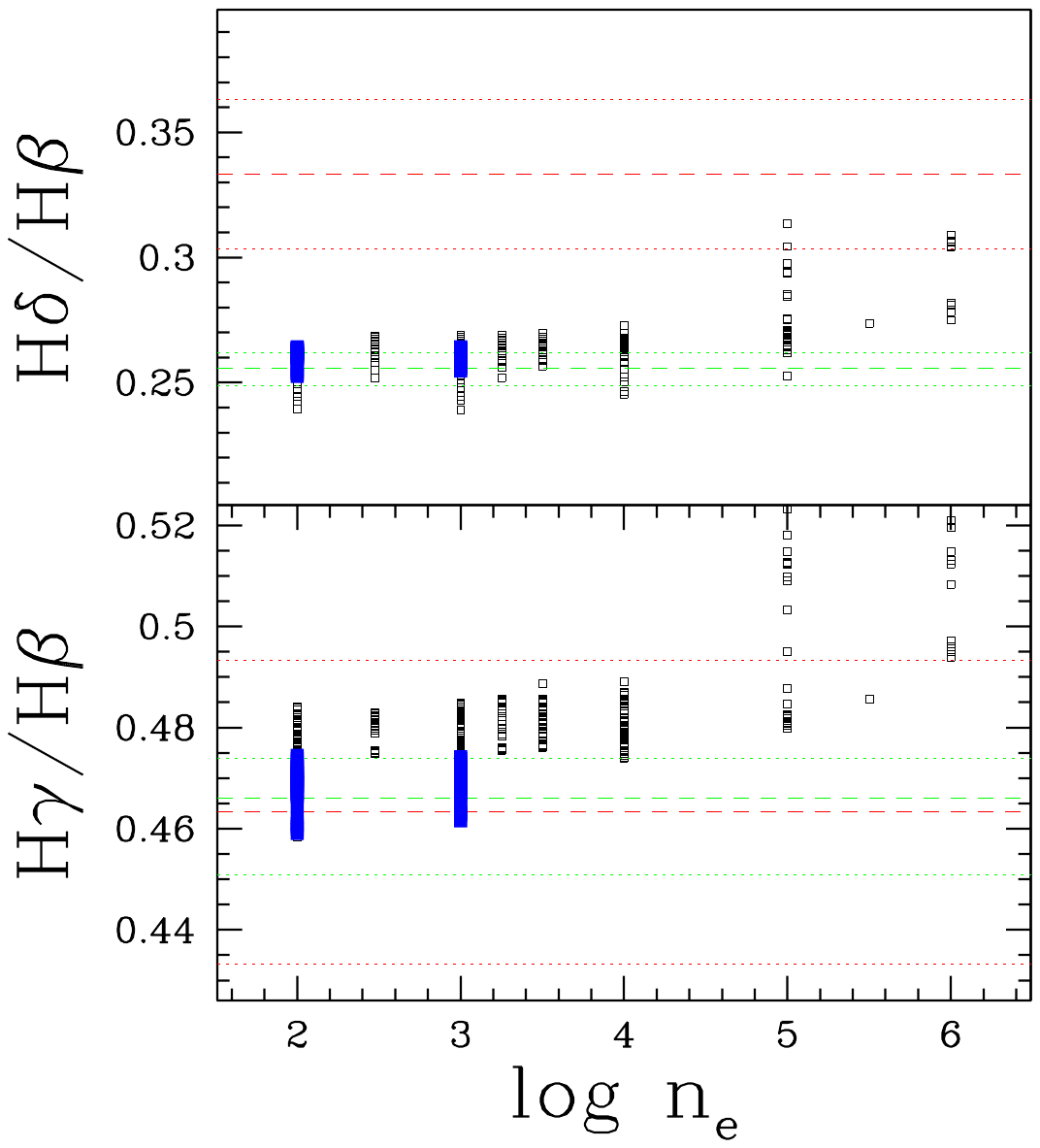}
	\caption{
	Ratio of ${\rm H\gamma/H\beta}$ (lower panel) and ${\rm H\delta/H\beta}$ (upper panel) as a function of the electron density. The red dashed line shows the measured value of the ratio for JD1 with the dotted lines showing the 1$\sigma$ error bars. The green dashed line shows the Case B value of the ratio for $10^4$ K with the dotted lines showing the $5\times 10^3$ K and $2\times 10^4$ K Case B values. Black squares represent a multitude of Cloudy models with varying properties, while the blue squares show the values for the ratio in \citet{Gutkin2016} models. 
	}
\label{fig:balmer}
\end{figure}

In Figure \ref{fig:balmer}, we show the measured values of ${\rm H\gamma / H\beta}$ and ${\rm H\delta / H\beta}$ (with their 1-$\sigma$ error band as a function of the electron density $n_e$ for a number of Cloudy \citep{Ferland2017} models that we have run as well as for \citet{Gutkin2016} models. The value we measure is incompatible with the presence of dust for gas temperatures below $2 \times 10^4$\,K. The absence of significant amounts of dust is also in agreement with the non-detection of FIR continuum for this object \citep{Hashimoto2018,tokuoka2022}.

%%%%%%%%%%%%%%

An intriguing possibility arising from Figure \ref{fig:balmer} is that JD1 might be characterized by a higher value of the electron density than the typically assumed values in the range 100-1000. Unfortunately, our spectra cannot resolve the [OII] doublet and therefore do not allow us to directly probe the electron density. %However, the detection of JD1 by ALMA at 88$\micron$ allows us to carry out a sanity check.
As an alternative, we can derive an estimate for the electron density from the [OIII]$\lambda 88\micron / {\rm [OIII]}\lambda 5007$ ratio. This approach has been adopted by, e.g., \citet{Fujimoto2022} for the galaxy ERO~S04590 at $z=8.5$.
In order to compare the ALMA 88$\micron$ measurement with our [OIII]$\lambda\lambda4959,5007$ measurements, we need to apply a significant aperture correction which will entail a relatively large uncertainty. 
We have determined an aperture correction using two different methods. The first approach is to assume that the ALMA flux is distributed across an area similar to the NIRCam image of JD1, with which we have estimated the aperture correction factor in Sec.~\ref{sec:nirspec} (i.e. 2.36). 
%We then scale the continuum level we see in the spectra near the line wavelengths to the one inferred from the best-fit SED (Sec.~\ref{sec:sfh}) and find an aperture correction of 0.425. 
% We then compare the total NIRCam flux to the continuum we see in the spectra to find an aperture correction of 0.425. 
Alternatively, we can derive an aperture correction by comparing the SFR derived from ALMA [OIII]$\lambda 88 \micron$ to that derived from \hb\ (Sec.~\ref{sec:sfrate}), to find an aperture correction of 1.69. \footnote{A magnification gradient across JD1 would introduce a difference between the ALMA and the MIRSpec aperture. In this paper we ignore this effect.} In the following we will adopt the average of these two values and adopt as error their semi-difference, i.e., $2.02 \pm 0.33$.
%$0.509 \pm 0.084$.

\begin{figure}
\centering
	\includegraphics[trim={0 5.5cm 0 4.5cm},clip,width=0.55\textwidth]{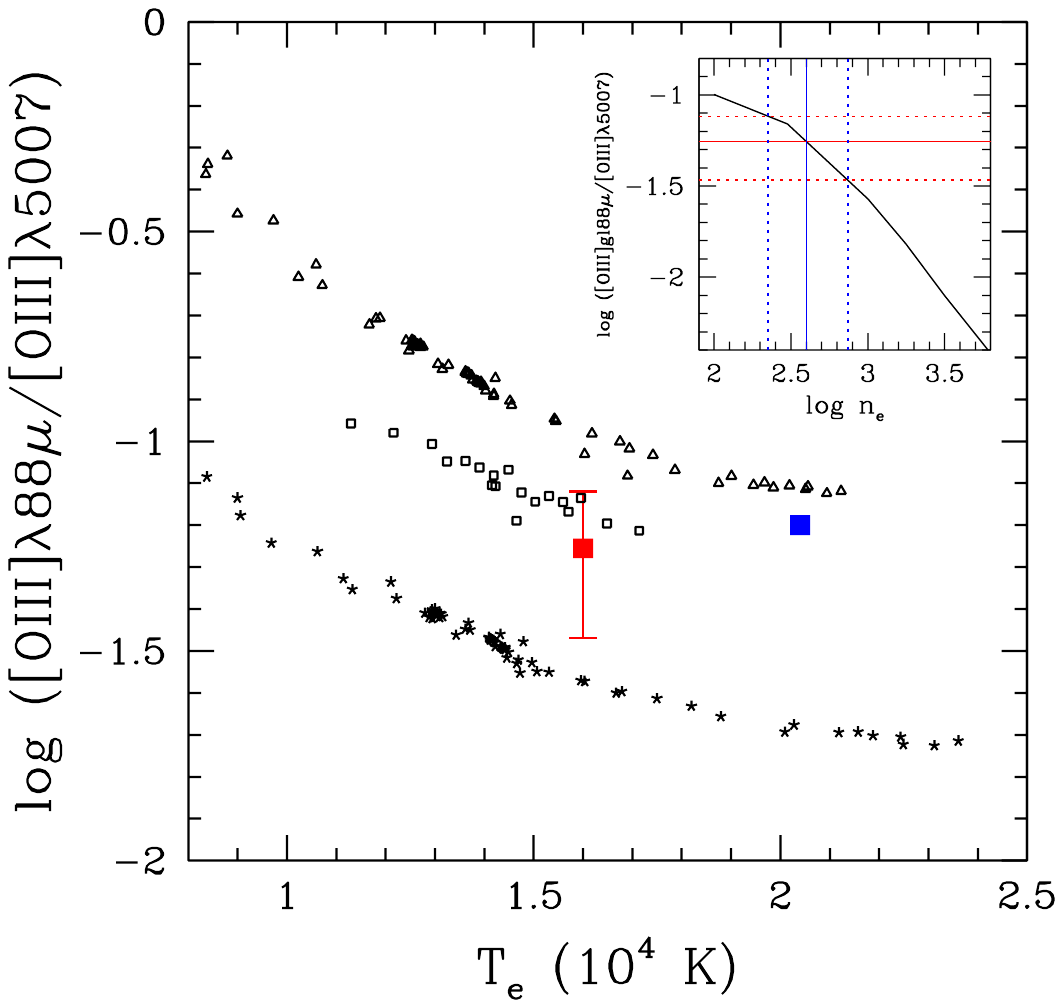}
	\caption{
	Ratio of $\log {\rm ([OIII]\lambda 88\micron / [OIII]\lambda 5007)}$ as a function of the electron temperature for a set of our Cloudy models. Triangles are for $n_e=100$, squares for $n_e=300$, stars for $n_e=1000$. The red square with an error bar is for JD1 and the blue square for ERO S04590. The inset shows the change of $\log {(\rm [OIII]\lambda 88\micron / [OIII]\lambda 5007)}$ as a function of electron density for an electron temperature of 16,000K. The red lines give the measured value for JD1 and its errors and the blue lines highlight the inferred values of the electron density.
	}
\label{fig:almaden}
\end{figure}

Armed with an estimate of the ALMA 88$\micron$ emission, we can derive the ${\rm [OIII]\lambda 88\micron / [OIII]\lambda\lambda}$4959,5007 ratio as $0.055 \pm 0.012 \pm 0.009$, where the first error is the measurement one and the second the aperture correction one. We adopt the same approach as \citet{Fujimoto2022} except by using Cloudy models instead of the PyNeb ones (which gives very similar results). 
Our Cloudy models \citep[see Section 3.2 of][]{oesch07} are based on a wide variety of electron densities, and use both a range of blackbody temperatures and a range of constant star formation rate of various metallicities modelled by GALAXEV \citep{BC2003}.
% Our Cloudy models are based on a wide variety of electron densities and use both a range of blackbody temperatures and a range of constant star formation rate Bruzual-Charlot models of various metallicities \citep{BC2003}.
Our results are shown in Figure \ref{fig:almaden}. For JD1 we find $\log n_e = 2.60^{+0.25}_{-0.27}$, or $n_e \simeq 400$. As a consistency check, for ERO S04590 we recover a value intermediate between 100 and 300 as found by \citet{Fujimoto2022}. 

On this basis, in the following we assume that JD1 has a density of $ n_e \simeq 400$ and no dust. 

%=============================

\section{Emission line analysis}\label{sec:emlines}

In the following, we will adopt the standard notation for line ratios, namely:

%    \vspace{0.4em}
%    {\small ${\rm R2 = \log ([OII]\lambda\lambda 3726,3729 / H\beta})$}
%    \vspace{0.4em}
    
%    {\small ${\rm R3 = \log ([OIII]\lambda 5007 / H\beta})$}
%    \vspace{0.4em}
    
    {\small ${\rm R23 = \log (([OIII]\lambda\lambda 4959, 5007 + [OII] \lambda \lambda 3726, 3729)/H\beta})$}
    \vspace{0.4em}
    
    {\small ${\rm O32 = \log ( [OIII]\lambda 5007 / [OII] \lambda \lambda 3726, 3729})$}
    \vspace{0.4em}
    
%    {\small ${\rm Ne3O2 = \log ( [NeIII] \lambda 3869 / [OII] \lambda \lambda 3726, 3729})$}
%    \vspace{0.4em}
    
With robust measurements of the auroral line [OIII]$\lambda$4363 and a value of the electron density, we can now apply the direct method to derive the electron temperature and the oxygen metallicity. The values derived at the two pointings are within the 1-$\sigma$ error bar and thus we will use the value from the combined spectrum. Using \citet{Aller1984} iterative method \citep[see also][]{Izotov2006} and $n_e = 400$, we derive $T_e({\rm O^{++}}) = 1.6 \times 10^4$\,K and infer $12 + \log {\rm O^{++}/H} = 7.88^{+0.04}_{-0.05}$. The error bars quoted here are those related to the measurement error. Varying the density between 100 and 1000 would contribute a 0.002 error. We have estimated $T_e({\rm O^{+}})$ using a variety of methods \citep{Izotov2006, Campbell1986, Pilyugin2016, Laseter2023, Sanders2023} and find values similar to or lower than $T_e({\rm O^{++}})$. Using these values we derive a contribution from ${\rm O^{+}}$ to metallicity $12 + \log {\rm O^{+}/H} = 6.39-6.55$ which is in any case small compared to the ${\rm O^{++}}$ contribution. For the total Oxygen metallicity, we adopt $12 + \log {\rm (O/H)}= 7.90^{+0.04}_{-0.05}$. We have also re-derived the direct metallicity for ERO S04590 and found values in agreement with the published ones.

%\begin{figure}
%\centering
%	\includegraphics[trim={0 6cm 0 0},clip,width=0.45\textwidth]{./fig/auroral.pdf}
%	\caption{
%	Ratio of $\log {\rm ([OIII]\lambda 4363 / [OIII]\lambda\lambda 4959,5007)}$ vs electron temperature $T_e$. The vertical blue line corresponds to the value of $T_e$ derived for JD1 assuming $n_e=400$. The solid red line is the measured JD1 value for the ratio (with the dotted line showing the 1$\sigma$ band). The points show the values for a set of Cloudy models. Triangles correspond to $n_e=100$, squares to $n_e=300$, stars to $n_e=1000$.
%	}
%\label{fig:auroral}
%\end{figure}

Using the methods outlined in \citet{Izotov2006} we also derive an estimate for the Neon abundance $\log ({\rm Ne/H}) = -5.03 \pm 0.04$. We correct the measurement with the Ionization Correction Factor (ICF) prescription by \citet{Amayo2021}, to find $\log ({\rm Ne/H}) = -5.02 \pm 0.04$ or $\log ({\rm Ne/O}) = -0.69$. Our measurements suggest that JD1 is under-abundant in Neon compared to solar by $ \Delta \log({\rm Ne/O}) \simeq -0.23$. 

We can also derive a Carbon over Oxygen ratio using the direct method \citep{ Aller1984,IzotovThuan1999}. Here we find $\log ({\rm C^{++}}/{\rm O^{++}}) = -1.09$. We can compute an ionization correction factor following \citet{Berg2019}. From the measured ${\rm O32} = 1.535 \pm 0.025$ and using the $Z=0.1 Z_\odot$ fit, we find $\log U = -1.647 \pm 0.026$ and a Carbon ICF of $1.216 \pm 0.013$. This gives us $\log ({\rm C/O}) = -1.08$, implying $\Delta \log({\rm C/O}) = -0.78$ or a C/O ratio of 0.16 of the solar value.

%As a sanity check, we show in Figure \ref{fig:auroral} the derived metallicity vs the measured ${\rm [OIII]\lambda4363/[OIII]\lambda5007}$ ratio for JD1 and compared the value derived from \citet{Izotov2006} iterative method (blue vertical line) with that from cloudy runs at different densities.

\begin{figure}
\centering
	\includegraphics[trim={0 6.5cm 0 4.5cm},clip,width=0.55\textwidth]{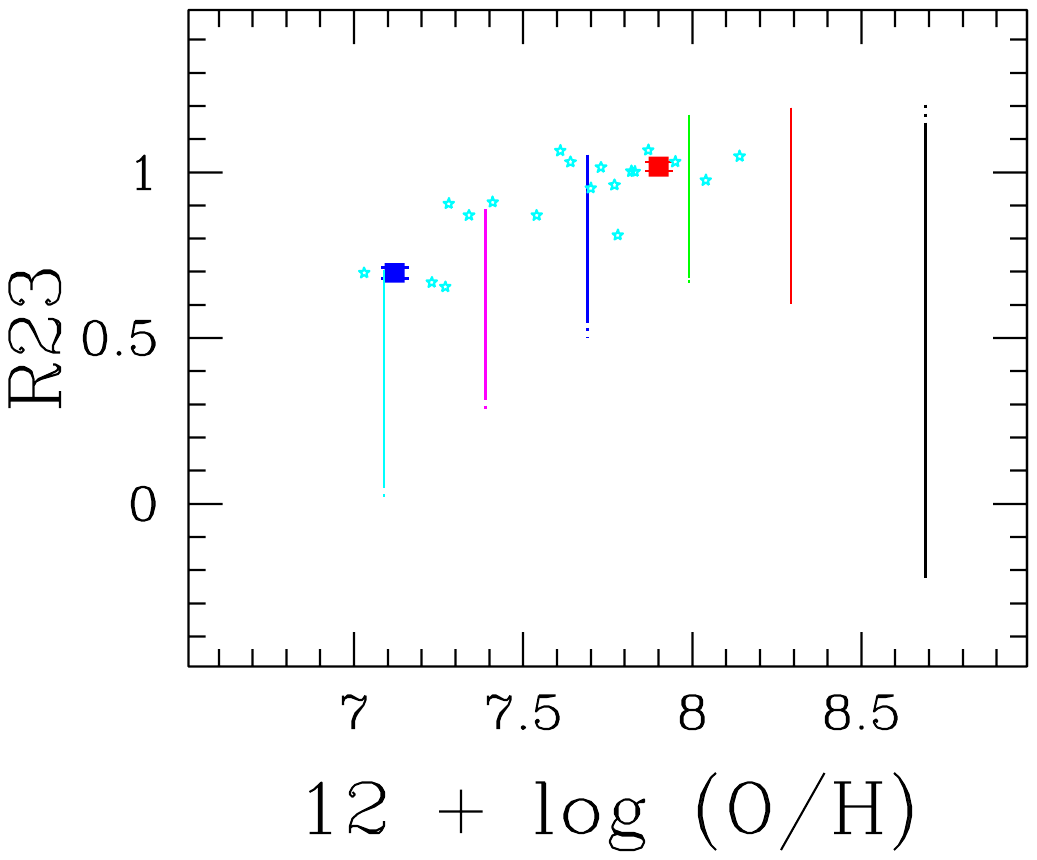}
	\caption{
	Direct oxygen metallicity measurement vs the $R_{23}$ ratio for JD1, ERO S04590 (blue) and other objects (cyan points) with auroral based measurements \citep{Nakajima2023, Laseter2023} For reference the vertical color lines are for \citet{Gutkin2016} models with increasing ionization parameter $U$ from bottom to top. 
	}
\label{fig:metals}
\end{figure}

Figure \ref{fig:metals} compares the metallicity and the R23 values we derive for JD1 with those derived for other auroral line measurements from the literature \citep{Laseter2023, Sanders2023}.

In order to explore strong line diagnostics, we considered \citet{Gutkin2016} models and found that the 
%In Figure \ref{fig:gutkin} we plot the measured line ratios for JD1 (red) and, for reference, ERO S04590 (in blue) compared to \citet{Gutkin2016} models. 
best fitting models correspond to 0.1 solar metallicity with $n_e=100$ and upper mass function cutoff $m_{\rm up} = 100 M_\odot$ and the highest values of the ionizing parameter explored in the models, $U = -1.5/-1$. This is in broad agreement with the direct measurement of the Oxygen metallicity.  The high level of ionization is also hinted at by the very high ${\rm [OIII]\lambda 5007 / H\beta} = 10.03 \pm 0.31$ which would correspond to the AGN-star formation boundary in the BPT \citep{BPT1981, Lamareille2009, Lamareille2010} diagram. Indeed, this diagram appears insufficient to separate star formation from AGN photoionization at   $z\geq6$ \citep{Cameron2023,Uebler2023}. However, we do not see any evidence of a broad component in the Balmer lines. 

The measurement of (C/O) for JD1 is consistent with Gutkin's (C/O) $\sim0.38$ solar models. Thus, the broad conclusions of the direct measurements, namely that JD1 has sub-solar metallicity, high-ionization, and Carbon-underabundance, could be derived from a strong emission line analysis, even though the details might not be precisely the same.

%\begin{figure}
%\centering
%	\includegraphics[trim={0 4.5cm 0 0},clip,width=0.45\textwidth]{./fig/carbonCGutkin.pdf}
%	\caption{We are showing the measured line ratios for JD1 (red) and, for reference, ERO S04590 (in blue) compared to \citet{Gutkin2016}  models. Black lines correspond to solar metallicity, red to 0.4 solar, green to 0.2 solar, blue to 0.1 solar, magenta to 0.05 solar, and cyan 0.025 solar. All models are for $n_e=100$ and $m_{up}=100 M_\odot$. The solid lines are for a solar [C/O] ratio. The dashed lines are for [C/O] = 0.38 the solar value and the dotted lines for [C/O] equal to 0.72 the solar ratio.  From the top left and clockwise we show: R3 vs $[CIII]\lambda 1909 / H\beta$, O32 vs R32, $\log ([OIII]\lambda 4363/H\gamma)$ vs R32, Ne3O2 vs R32.  Solid lines are for solar abundance ratios, dotted lines for 0.72 (C/O) and dashed lines for 0.38 (C/O). For clarity, in the R3 vs $[CIII]\lambda 1909 / H\beta$ panel (upper left) we use thin lines for metallicities other than 0.1 solar.
% }
%\label{fig:gutkin}
%\end{figure}

\section{On the Non-solar abundance ratio}\label{sec:COratio}

We have seen that the direct method indicated that JD1 has C/O ratio of 0.16 the solar value. We should note the large difference in wavelength between CIII]$\lambda1909$ and \hb, and the fact that they fall on different NIRSpec gratings. 
The measurement of (C/O) depends also on an estimation of an ionization correction based on the O32 ratio which is itself vulnerable to the presence of dust. This opens up the possibility of uncertainties due to dust (i.e. an additional correction that we have not applied) and uncertainties in the instrumental calibration. The latter, however, should be less than 5\,\%. Acknowledging this potential source of uncertainty, we now proceed to examine the implication of the derived under-abundance of Carbon.

%As a consistency check, we can verify whether \citet{Gutkin2016} support the evidence from the direct measurement. In Figure \ref{fig:carbon} we show the expected value of $Log [CIII]\lambda 1909 / H\beta$ for \citet{Gutkin2016} models compared to the value we measure. The solar (C/O) ratio (solid line) is strongly disfavored but $(C/O) = 0.38 (C/O)_\odot$ goes through the JD1 point.  Unfortunately the more extreme under-abundance found with the direct method is not supported by this comparison. At face value, this result confirms that JD1 is Carbon-poor compared to the solar value but how under-abundant might need to be revisited. However, in the following, we will continue to adopt the direct value (C/O) ratio.

A similar under-abundance has been found in a galaxy at $z=6.23$ by \citet{Jones2023}. However, that galaxy has an overall lower metallicity than JD1, which may make a very young age and enrichment dominated by core-collapse supernovae less surprising.
Comparing JD1 with the results by \citet{Arellano2022}, looking at the top panels of their Figure 4, one notices that JD1's (C/O) is lower than what they find for their $z>7$ galaxies but is not anomalous compared to $z \sim 2$ galaxies. The same is true for the (Ne/O) value of JD1 is not observed locally for objects of its metallicity.

The under-abundance of Carbon exceeds the under-abundance of Neon (see Section \ref{sec:emlines}). It should be noted that while a low C/O is a common feature of core-collapse supernova yields \citep[e.g.,][]{Nomoto2013}, the same is not true for (Ne/O), even though there are models where this is the case \citep[e.g.,][model S20]{Rauscher2002}. Massive, Population III,  PISN supernovae can also lead to significantly under-abundant (Ne/O) \citep{Heger2010}.

%\begin{figure}
%\centering
%	\includegraphics[trim={0 15cm 0 0},clip,width=0.45\textwidth]{./fig/carbonGutkin.pdf}
%	\caption{We are showing the measured line ratios $Log [CIII]\lambda 1909 / H\beta$ for JD1 (red) vs R3 (left) and Ne3O2 (right).  We also show for comparison \citet{Gutkin2016}  models for 0.1 solar abundance. All models are for $n_e=100$ and $m_{up}=100 M_\odot$.   The solid lines are for a solar [C/O] ratio. The short dashed lines are for [C/O] = 0.38 the solar value, long dashed lines are for [C/O]= 0.20. and the dotted lines for [C/O] equal to 0.72 .}
%\label{fig:carbon}
%\end{figure}

%=============================
\section{Star formation history and absence of old stellar populations}\label{sec:sfh}

The sensitive photometry over 1--5\,$\mu$m by NIRCam enables us to constrain the stellar component in JD1. The F356W and F444W data points are critical, as they cover, for the redshift of JD1, the rest-frame $3800$--$4200$\,\AA\ where the Balmer break, a characteristic break for relatively older (i.e. B, A, and F-type) stars, is located. Previous studies using Spitzer Ch1 and Ch2 reported red color \citep[$m_{\rm ch1}-m_{\rm ch2}\sim0.9$\,mag;][]{huang16,zheng17,hoag18}, speculating the presence of such old populations formed at $z\sim15$ \citep[][]{Hashimoto2018}. Here, using our JWST data we report $m_{356}=25.76\pm0.02$ and $m_{444}=25.63\pm0.02$ i.e. the absence of such characteristic features in JD1. As the F356W magnitude is consistent with the previous IRAC Ch1 measurements, we suspect that image confusion might have affected the Spitzer IRAC Ch2 measurements.

The NIRCam photometry reveals a blue color, with $m_{\rm F356W} - m_{\rm F444W} = 0.13\pm 0.01$\,mag. This is considerably smaller than the previous measurement. We note that the filter curves of the corresponding bands of the two telescopes are only slightly different. For the redshift of JD1, \oiii $\lambda$5007 line is within the red edge of F444W, but a hair outside the Spitzer Ch2. As such, the absence of strong color in F356W/F444W (despite the strong \oiii\ included in the redder filter) indicates the dominance of young ($<100$\,Myr) stellar populations.

Our spectrum is deep enough to reveal the continuum, and thus allows us to directly measure the Balmer break. Following the definition of \citet{Balogh1999}, we measure $D(4000)\,\sim0.5$, meaning that the continuum at the blue side is brighter than the red side. We also follow the procedure presented in \citet{curtis-lake23} and measure the strength of the Balmer break $0.5\pm0.1$, supporting the dominance of young populations.

Lastly, we perform a joint SED analysis combining photometric and spectroscopic data using {\tt gsf} \citep{morishita19}. Briefly, we use {\tt fsps} \citep{conroy09fsps}, with the default MIST isochrone \citep{choi16} and MILES stellar library \citep{Falcon-Barroso11}. Both stellar- and gas-phase metallicity are fixed to 1/5 the solar, and dust is set to zero ($A_V=0$), based on the results from our ISM analyses above. %are set as free parameters, while the metallicity of nebular component is fixed during the fit to the same one of the corresponding stellar component.{\bfr <-- is this what you meant?} 
We adopt a binned star formation history, which offers flexible inference on star formation history, with a set of ages [0.001,0.003,0.01,0.03,0.1,0.3,0.5]\,/\,Gyr. We find that JD1 experienced two distinct, major star formation phases in the past --- one $\sim300$\,Myr old ($z\sim14$) and a more intense one 10\,Myr old. The presence of a more recent burst is consistent with the Carbon deficit that we have found in our spectral line analysis. We note that the absence of a significant amount of old stars results in a much smaller stellar mass in JD1, by a factor of $\sim\times6.8$ from previous measurements, namely a stellar mass of $1.6_{-0.3}^{+0.3} \times 10^8 (10/\mu) {\rm M}_\odot$. Combined with the SFR from Section \ref{sec:sfrate} we find a specific star formation rate of $37_{-7}^{+10} {\rm Gyr}^{-1}$.

 To check for consistency among various assumptions in SED fitting, we perform another SED analysis with {\tt BAGPIPES} \citep{Carnall2018, Hsiao2023} using {\tt BPASS} v2.2.1 stellar population models \citep{Stanway2018} and {\tt CLOUDY} c17.03 photoionisation code \citep{Ferland2017}. We also fix the metallicities to 0.15 solar and dust to null. We assume a smooth, Gaussian process-based nonparametric SFH \citep{Iyer2019}. We set the SFH to be controlled by four parameters: stellar mass, SFR, and two shape parameters, which essentially divide the SFH into three lookback time intervals in which the galaxy formed equal mass. The derived SFH agrees reasonably well with the binned SFH from above. The resulting stellar mass is $1.42_{-0.04}^{+0.05} \times 10^8 (10/\mu) {\rm M}_\odot$. This stellar mass gives us a specific star formation rate of $40_{-2}^{+3} {\rm Gyr}^{-1}$.
%The SFR is $9.4_{-0.4}^{+0.3} (10/\mu) {\rm M}_\odot/\rm{yr}$ while the log(sSFR) is $-8.18_{-0.02}^{+0.02}.

\begin{deluxetable}{lr}
    \tablecolumns{2}
    \tablecaption{
    Observed properties of JD1.
    }
    \tablehead{
    \colhead{} & \colhead{}
    }
    \startdata
    \cutinhead{Redshift and Line Flux Measurements}
    $z^{\dagger}$ & $9.114\pm0.001$\\
    ${\rm H\beta_{\lambda4861}}$ & $375\pm 10$\\
    ${\rm H\gamma_{\lambda4340}}$ & $174\pm 6$\\
    ${\rm H\delta_{\lambda4102}}$ & $125\pm 6$\\
    ${\rm [OII]_{\lambda\lambda3727,3729}}$ & $110\pm 5$\\
    ${\rm [OIII]_{\lambda4363}}$ & $59\pm 5$\\
    ${\rm [OIII]_{\lambda\lambda4959,5007}}$ & $3764\pm 14$\\
    ${\rm [NeIII]_{\lambda3869}}$ & $154\pm 5$\\
    ${\rm CIII]_{\lambda\lambda1907,1909}}$ & $139\pm 8$\\
    \cutinhead{Physical properties}
    $M_{\rm UV} / {\rm mag} $ & $-19.33_{-0.01}^{+0.01} - 2.5 \log(10/\mu)$\\
      SFR$_{\rm H\beta} / M_\odot\,{\rm yr^{-1}}$ & $5.9\pm 0.2\,(10/\mu)$\\
    SFR$_{\rm SED} / M_\odot\,{\rm yr^{-1}}$ & $3.37_{-0.07}^{+0.07}\,(10/\mu)$\\
    $M_*/10^8\,M_\odot$ & $1.61_{-0.25}^{+0.25} (10/\mu)$\\
    $\beta_{\rm UV}$ & $-2.20_{-0.01}^{+0.01}$\\
    $A_V / {\rm mag}$ & [0.0]\\
    $Z / {Z_\odot}$ & [0.2]\\
    \enddata
    \tablecomments{
    $\dagger$: Redshift measured for the \hb+\oiii$_{\lambda\lambda4959,5007}$ lines. Flux measurements are corrected for aperture loss and expressed in units of $10^{-21}\,(10/\mu)$\,erg\,s$^{-1}$\,cm$^{-2}$.
    }\label{tab:prop}
    \end{deluxetable}

%=============================
\section{Discussion and Summary}\label{sec:disc}

We find that JD1 is an actively star forming galaxy  characterized by metallicity about 1/5 the solar value and with emission line ratios best explained by a very strong ionizing continuum suggestive of a very young stellar population.  The youth of the object is also supported by the star formation history we derived, showing a major burst $\sim10$\,Myrs ago, and by our finding of a low (C/O) ratio suggestive of Oxygen enrichment by supernovae, while intermediate mass stars have not yet had the time to increase the Carbon fraction. Such indications of a very active, recent and on-going star formation are somewhat surprising when coupled with the lack of evidence for dust as indicated by the measured Balmer decrement as well as the non-detection in the ALMA continuum reported in the literature. While geometry could conspire to give rise to grey dust \citep[see e.g.][]{Witt1996, Witt2000}, it is interesting that this object seems also qualitatively compatible with the dust ejection scenario described by \citet[][see also \citealt{Tsuna2023}]{Ferrara2023}. Indeed, the value of specific star formation rate we measure  exceeds the value of $\sim 32 {\rm Gyr}^{-1}$  that we derive from Eq. 3 of \citet{Ferrara2023}. It should be noted that the complexities of how supernova driven winds would affect the observed C/O ratio would need to be evaluated in more detail \citep[see, e.g.,][]{Berg2019}.

It is worth noticing that the (C/O) ratio of JD1, as well as its metallicity and high ionization (as inferred from the O32 ratio) are comparable to what is observed in some Green Peas galaxies at $z<0.3$ \citep{Swara2020,Rhoads2023}.

We have seen in Section \ref{sec:nircam} that JD1 is characterized by multiple components. This could be an indication of recent interaction or merger. However, we see no evidence of color gradients, and the line ratios at the two slightly offset slit positions are compatible. Given the absence of a color gradient, it is hard to assess how general our conclusions can be once one opens up geometry and substructure. For completeness we have also considered two component Cloudy models. The significant number of additional degrees of freedom prevents us from obtaining a unique solution but our best two-component fits share similar properties of hard ionizing flux, sometimes with AGN or an additional stellar population with $8 \times 10^4$\,K of effective temperature, and oxygen enriched models. Given the high level of degeneracy, we do not consider it useful to discuss these solutions in greater detail except to say that they are broadly compatible with the conclusions of single component modeling.

While completing this paper we were sent a pre-publication copy of a paper on JD1 by Brada{\v{c}} et al., we find their results broadly compatible with ours given the differences in JWST data for the two papers. Furthermore, JD1 has been observed using the NIRSpec IFU by another program (PID: 1262, PI: N. Luetzgendorf) but we have currently no access to these data.  It will be interesting to see whether the spatial resolved spectroscopy modifies the picture presented here and how a combined analysis of both data and of our Cycle~3 observations (\#4552) aiming at observing with high resolution the rest frame UV of JD1 will affect the picture.

%%%%%%%%%%%%%%%%%%%%%

%=============================

%=============================
\section*{Acknowledgements}
This work is based in part on observations made with the NASA/ESA/CSA James Webb Space Telescope. The data were obtained from the Mikulski Archive for Space Telescopes at the Space Telescope Science Institute, which is operated by the Association of Universities for Research in Astronomy, Inc., under NASA contract NAS 5-03127 for JWST. The specific observations analyzed can be accessed via \dataset[DOI: 10.17909/2q9q-mw78]{https://doi.org/10.17909/2q9q-mw78}. These observations are associated with program JWST-GTO1199. Support for program JWST-GTO1199 was provided by NASA through grant 80NSSC21K1294. We thank S. Charlot for sharing a broader set of emission lines for the \citet{Gutkin2016} models. The authors wish to thank M. Brada{\v c}, R. Ellis, A. Ferrara, T. Hashimoto, C. Leitherer, R. Maiolino,  G. Roberts-Borsani,  S. Suyu for discussions.  We thank the anonymous referee for comments that helped improve the paper. 

%{{\it Software:} Astropy \citep{astropy:2013, astropy:2018}, numpy \citep{oliphant2006guide,van2011numpy}, python-fsps \citep{foreman14}, EMCEE \citep{foreman13}. }


\bibliography{sample631}{}
\bibliographystyle{aasjournal}



\end{document}